\documentstyle[preprint,aps]{revtex}
\begin{document}
\draft
\preprint{\vbox{\hbox{MRI-PHY/19-95}\hbox{IASSNS-HEP-95-101}
\hbox{IP-BBSR/95-87}}}

\baselineskip= 18 truept

\def\w{\wedge}
\def\e{\epsilon}
\def\ra{\rightarrow}
\def\a{\alpha}
\def\b{\beta}
\def\n{\eta}
\def\g{\gamma}
\def\d{\delta}
\def\t{\theta}
\def\o{\Omega}
\def\p{\phi}
\def\P{\Phi}
\def\cg{\cal G}
\def\cb{\cal B}
\def\pr{\partial }
\def\bpr{\bar {\pr }}
\def\ca{{\cal A}}
\def\cl{{\cal L}}
\def\cm{{\cal M}}
\def\zb {{\bar z}}
\def\na{{\nabla }}
\def\s{{\sigma}}
\def\G{{\Gamma }}
\def\l{{\lambda }}
\newcommand{\be}{\begin{equation}} \newcommand{\ee}{\end{equation}}
\newcommand{\bea}{\begin{eqnarray}}\newcommand{\eea}{\end{eqnarray}}

\title{\bf Dualities in Five Dimensions and Charged String Solutions }

\author{{\large{Supriya Kar$^1$, Jnanadeva Maharana$^{2\dag }$
and Sudhakar Panda$^1$}}
\footnote{{e-mail: supriya/panda@mri.ernet.in}\hfil\\
{$^{\dag }$Permanent address: Institute of Physics, Bhubaneswar, India;
e-mail: maharana@sns.ias.edu}}}

\address{{\vspace{.1in}}
{$^1$Mehta Research Institute of Mathematics
$\&$ Mathematical Physics\\
10, Kasturba Gandhi Marg, Allahabad-211 002, India}}

\address{{\vspace{.3in}}
{$^2$School of Natural Sciences, Institute for Advanced Study\\
Olden lane, Princeton NJ 08540, USA}}


\maketitle

\thispagestyle{empty}

\begin{abstract}
\baselineskip= 14 truept

We consider an eleven dimensional supergravity compactified on
$K3\times T^2$ and show that the resulting five dimensional theory has
identical massless states as that of heterotic string compactified on a
specific five torus $T^5$. The strong-weak coupling duality of the five
dimensional theory is argued to represent a ten dimensional Type $IIA$
string compactified on $K3 \times S^1$, supporting the conjecture of
string-string duality in six dimensions. In this perspective, we present
magnetically charged solution of the low energy heterotic string
effective action in five dimensions with a charge defined on a three
sphere $S^3$ due to the two form potential. We use the Poincare duality to
replace the antisymmetric two form with a gauge field in the effective
action and obtain a string solution with charge on a two sphere $S^2$
instead of that on a three sphere $S^3$ in the five dimensional spacetime.
We note that the string-particle duality is accompanied by a change of
topology from $S^3$ to $S^2$ and viceversa.

\end{abstract}

\vfil
\newpage

\section{Introduction}

Our understanding of string theory is strongly guided by its
rich symmetry contents. Among various symmetries, the target space
symmetries such as $T$-duality and $S$-duality have been at the center of
attention in the recent years. It is noted that the former can be tested
in the perturbative frame work, interchanging the Kaluza-Klein modes with
the string winding modes, whereas the latter is a nonperturbative
phenomena and relates the strong and weak coupling regimes.

\vspace{.1in}
In this context, we briefly review some of the essential evidences related
to various dualities. There have been some progress to understand
the nonperturbative aspects of string theory\cite{1} by studying the
string-fivebrane duality\cite{2} in ten dimensions. It has been
argued that the fundamental string and the solitonic string are two
different formulations of a single theory with strongly coupled string
corresponding to weakly coupled fivebrane and viceversa\cite{3}. For
example, if $e^{\p_0}$ is the string coupling ($\p_0$ being the dilaton)
then the corresponding fivebrane coupling is $e^{{-\p_0}/3}$. A nice
review can be found in ref.\cite{4}.

\vspace{.1in}
Recently, the dynamics of string in various dimensions have been discussed
by Witten \cite{5}, suggesting a series of interconnections between
different string theories. There are evidences\cite{6} that
the six dimensional heterotic string derived from the ten dimensional one
after compactification on $T^4$, is dual to Type $IIA$ string in six
dimensions which is obtained from its ten dimensional counterpart by
compactification on $K3$. As a result, the spectrum of Bogomol'nyi
saturated states in the heterotic string theory on $M^6\times T^4$ is
identical to that of Type $IIA$ string on $M^6\times K3$\cite{7}. It has
been shown that the string-string duality in six dimensions relates the
fundamental heterotic string singular solutions to the solitonic heterotic
string\cite{8}. In this context, it has been conjectured that an eleven
dimensional supergravity theory compactified on $K3$ can be interpreted as
a heterotic string theory on a three torus following string-membrane
duality\cite{9} in seven dimensions. At present with the exisiting duality
conjectures, it is argued that the eleven dimensional supergravity theory
with an underlying low energy membrane theory is a plausible candidate for
the origin of various string theories\cite{10}. In this connection, one of
the authors  has considered\cite{11} the Type $IIA$ and Type $IIB$
string theories in eight dimensions and has shown that the resulting
p-branes can be classified in the $SL(3,Z)\times SL(2,Z)$ representaion.
The strong reason behind all the speculations is the fact that the
resulting theories have identical massless moduli spaces of vacua and the
low energy spectrum along with same supersymmetry.

\vspace{.1in}
Now, let us recall the idea of duality between electric and
magnetic charges in four dimensions\cite{12}, in connection to the
monopole solutions\cite{13} to Yang-Mills-Higgs theory. Subsequently, the
electric-magnetic duality under the name $S$-duality have been discussed
in $N=4$ supersymmetric Yang-Mills theory\cite{14}. However, at present
there are increasing evidences for the $S$-duality symmetries relating the
strong and weak coupling regimes in the $N=2$ supersymmetric gauge
theories\cite{15} and in $N=1$ supersymmetric theories\cite{16}. In
sequel, the duality symmetry in string theories\cite{17} have been studied
in great detail in order to understand the string theory
nonperturbatively\cite{18}. The stringy realization of the results of
$N=2$ Yang-Mills theory\cite{15} have been discussed in
literature\cite{19}. It has been shown that the duality transformation
exchanges the electrically charged elementary string excitations with the
magnetically charged soliton states\cite{20} in the Narain's toroidal
compactification\cite{21} of heterotic string theory. There are also
attempts to describe fundamental string as solitons in the dual string
theory by utilizing various analogies between the superstrings with the
solitons in the supersymmetric theories\cite{22}. These string solitons
can be constructed from the fivebrane solitons\cite{23} with four of their
world volume dimensions wrapped in the internal directions. In
ref.\cite{gibb} instanton and brane solutions are discussed in Type $IIB$
string theory. Some of these solitons have been interpreted as the
monopole solutions\cite{24}. It has been shown that they saturate the
Bogomol'nyi bound preserving half of the supersymmetries\cite{25}. These
$BPS$ saturated states may play an important role in the nonperturbative
dynamics of string theory. It has been argued that the study of such
configurations may shed light on duality between strong-weak coupled
string vacuua.

\vspace{.1in}
Furthermore, in a series of papers\cite{5,26,27,28,29,30,31,berg}, various
predications of the duality conjectures are tested for the heterotic
string theory along with Type $IIA$ and $IIB$ string theories. One of the
most exciting result is that the strong coupling limit of Type $IIA$
string in ten dimensions is an eleven dimensional
supergravity\cite{5,32}, representing an underlying
low energy membrane theory\cite{33}. Since a membrane is $T$-dual to a
fivebrane theory in eleven dimensions, the string-fivebrane duality in ten
dimensions can be interpreted as a result of double dimensional reduction
on $S^1$ of the corresponding membrane-fivebrane duality in eleven
dimensions\cite{34}. However, the membrane and fivebrane theories have not
been understood completely as the quantum corrections are yet to be
discovered. Note that, whenever we refer to the eleven dimensional
supergravity theory, the underlying low energy membrane theory of world
volume is understood.

\vspace{.1in}
In this paper, we investigate the dynamics in five dimensions
corresponding to dimensionally reduced membrane theory, heterotic string,
Type $IIA$ and Type $IIB$ string theories with $N=2$ supersymmetry. We
seek classical solutions of the heterotic string effective action with a
motivation to provide evidence for various dualities. Note that, in five
dimensions the two form potential is Poincare dual to the gauge field and
the elementary string states are not charged in the spectrum with respect
to this new gauge field. It has already been argued that such an electric
charge corresponding to the new gauge field may arise in order to satisfy
the anomaly equation involving three rank antisymmetric field strength and
gauge field strength\cite{5}. We show that the string-particle duality in
five dimensions leads to various theories admitting classical solutions
with charge defined on different toplogies, $e.g.\; S^3$ and $S^2$
etc. We present a five dimensional solitonic magnetically charged string
solutions of the heterotic string effective action and apply Poincare
duality to obtain another charged solution with charge defined on a
different topology than the one before. The important point specific to
five dimensions is the fact that the mass of charged $BPS$ states is
inversely proportional to the square of the coupling constant of the
theory. Thus, in the strong coupling limit the mass of the multiplets tend
to zero which is very unlikely in order to interpret the mass spectrum in
five dimensions. Similarly, in the weak coupling limit, the mass of the
charged states diverges. Thus the $BPS$ value for the mass of the
charged states leads to difficulty in interpreting the mass spectrum in
the strong and weak coupling regimes. However it is argued that these
charged states can be interpreted as Kaluza-Klein states on $R^5\times
S^1$ in order to overcome the puzzle in five dimensions\cite{5}. The
$T-$duality does not act on the new compactified coordinate represented by
a circle $S^1$. The above analysis in five dimensional theories has
pointed out the need for a six dimensional theory (as the radius of $S^1$
is very large) in the strong coupling regime. As a result, the
uncompactified theory seems to have eleven dimensions instead of ten
dimensions. Thus it is natural to start with an eleven dimensional
supergravity theory\cite{35} which has an underlying low energy membrane
interpretation.

\vspace{.1in}
Now, we propose to explore the duality symmetry between ten dimensional
heterotic string theory compactified on a five dimensional torus $T^5$, a
Type $IIA$ string on $K3\times S^1$, Type $IIB$ string on $K3\times S^1$
and an eleven dimensional $N=1$ supergravity theory compactified on
$K3\times T^2$. The bosonic part of the ten dimensional heterotic string
action contains massless fields such as dilaton, graviton, antisymmetric
tensor field and $16$ abelian gauge fields. We show that the
resulting five dimensional heterotic string theory has $27$ gauge bosons;
$26$ of which transform as vectors under the global noncompact group
$O(5,21)$ and one comes after Poincare dualizing the field strength of
antisymmetric tensor. On the other hand, compactification of the eleven
dimensional $N=1$ supergravity theory on $K3\times T^2$ has
$N=2$ spacetime supersymmetry and describes a five dimensional low energy
string effective action. At generic points in the moduli space of
$K3\times T^2$, the gauge symmetry of $D=11$ supergravity theory is
enhanced similar to the case of Type $IIB$ superstring compactified on the
$K3\times S^1$ \cite{5} and Type $IIA$ string compactified on
$K3$\cite{36}. It has been argued\cite{5} that for a Type $IIB$ string
compactified on $K3$, the $U$-duality group is $SO(21,5;Z)$; which is a
combination of the corresponding $T$-duality group $SO(20,4;Z)$ and the
$SL(2,Z)$ symmetry of the ten dimensional Type $IIB$ string theory.
In fact the moduli space of vacua of Type $IIB$ theory compactified on
$K3$ is identical to that of a heterotic string theory in five dimensions.
However there are 5 self dual two forms, 21 antiself dual two forms in
contrast to the 26 gauge fields in case of the heterotic string theory in
five dimensions. Further compactification of the six dimensional Type
$IIB$ string on a circle $S^1$ gives rise to 26 gauge fields with enhanced
gauge symmetry at the generic points in the moduli space and can be
identified with a heterotic string compactified on a five torus.

\vspace{.1in}
We outline the paper as follows. In section II, we present a low energy
heterotic string effective action in five dimensions obtained by toroidal
compactification of the ten dimensional one and discuss the background
field contents. The Poincare duality in $D=5$ is used to construct the
dual form of the effective action. In section III, we deal with the eleven
dimensional supergravity which can be considered as the low energy limit
of a membrane. We write down the $D=11$ supergravity theory compactified
on $K3$ and further compactify it on a two torus to obtain a five
dimensional theory. In fact we use double dimensional reduction on $K3$ to
arrive at the underlying worldsheet picture in seven dimensions from the
corresponding worldvolume describing a membrane in eleven dimensions. In
the next step, we compactify the spacetime on a two torus $T^2$ and
arrive at a five dimensional theory possessing string-particle
duality from the original eleven dimensional membrane-fivebrane duality.
With field redefinitions along with the Poincare duality on
the three form potential, we show in section IV that the number of moduli
fields and the gauge field multiplets are identical to that of a
toroidally compactified heterotic string in $D=5$. The strong-weak
coupling limit of eleven dimensional supergravity theory compcatified on
$K3\times T^2$ is discussed in section V and the massless spectrum is
identified with that of Type $IIA$ string compactified on a $K3\times
S^1$, supporting the conjecture in six dimensions. In section VI, we
present solitonic solution of the heterotic string effective
action with charge defined on a three sphere $S^3$ due to the two form
potential. We analyze another magnetically charged solution with a $S^2$
geometry of the dual heterotic string effective action and
discuss the consequences of the Poincare and strong-weak couple dualities
under the name string-particle duality in five dimensions. Finally in
section VII, we summarize our result with various interconnections between
different string theories in five dimensions along with the classical
charged solutions.

\section{Heterotic string effective action in five dimensions}

We begin this section by toroidally compactifying a $D=10$
heterotic string theory to a five dimensional one. The resulting theory
has $N=2$ supersymmetry and contains five Kaluza-Klein gauge fields and
five winding gauge fields from the ten dimensional metric and
antisymmetric tensor respectively. The elementary string states are
electrically charged with respect to these gauge fields. In addition to
the ten gauge fields there is also one more spin one field which is
indeed Poincare dual to the antisymmetric tensor in five dimensions.
However the state corresponding to the new gauge field is not
necessarily electrically charged because of its appearance specific to
five dimensions. The five dimensional effective action at generic points
in the moduli space is manifestly invariant under the $O(5,21; Z)$
transformations.

\vspace{.1in}
Now we start by writing down the bosonic part of the low energy heterotic
string effective action compactified on a five torus $T^5$ following the
general prescription of ref.\cite{37} as

\bea
S_h= \int d^5x{\sqrt {-g}}e^{-\P}\Big [ R&+&g^{\mu\nu}\pr
_{\mu}\P \pr _{\nu} \P -{1\over12}
g^{\mu\mu '}g^{\nu\nu '}g^{\rho\rho '} H_{\mu\nu\rho}
H_{\mu '\nu '\rho ' }\nonumber\\
&-&{1\over4}g^{\mu\mu ' }g^{\nu\nu '}
F^{(a)}_{\mu\nu } (\cl\cm\cl)_{ab} F^{(b)}_{\mu '\nu ' } + {1\over8}
g^{\mu\nu }\ Tr(\pr _{\mu } \cm\cl\pr _{\nu }\cm\cl)\Big ],
\eea
where the three form antisymmetric field strength is obtained from the
ten dimensional counterpart by toroidal compactification and is defined
with a gauge Chern-Simon term as
$$H_{\mu\nu\rho } = {\pr_{[ \mu } B_{\nu\rho ]}}
-\ {1\over2}A^{(a)}_{[ \mu }{\cl}_{ab}F^{(b)}_{\nu\rho ]} $$
and the two form gauge field strengths are given by
\be
{F^{(a)}}_{\mu\nu } =\ \pr_{\mu }A^{(a)}_{\nu }\ -\ \pr_{\nu}A^{(a)}_{\mu }.
\ee
The three form $H$ satisfies the modified Bianchi identity involving the
gauge field strengths $F^{(a)}$\cite{5}. The massless bosonic background
field configuration  $(0\le \mu\le 4)$ include the metric $g_{\mu\nu}$,
the antisymmetric tensor $B_{\mu\nu}$, the abelian gauge fields
$A^{(a)}_{\mu}$ for $a=1,2\dots 26$, the dilaton $\P$ and a $(26\times
26)$ matrix valued scalar field $\cm $ representing an element of the
coset given by
\be
\cm\ \in {O(5,21)\over O(5)\times O(21)},
\ee
where $\cm ^T = \cm $ and $\cm\cl\cm ^T = \cl ,$ for
$\cl \ ={\pmatrix {{-I_5} & {0} \cr {0} & {I_{21}}\cr }}.$

\vspace{.1in}
\noindent The moduli field $\cm $ is constructed from the internal
components of a set of ten dimensional fields namely; the metric, the
antisymmetric tensor field and the gauge fields. There are $15$ scalars
originating from the ten dimensional metric, $10$ of them from ten
dimensional antisymmetric tensor and the remaining $80$ of them from the
gauge fields in ten dimensions. Altogether, there are $106$ scalar
fields out of which $105$ of them ($\varphi ^i$) arise from the
parametrization of the coset space (3) in terms of target space
coordinates, $i\ =\ 1,\ 2,\ ...\ 105$ and the remaining one represents a
dilaton $\P$. There are altogether $26$ gauge fields, out of which
$5$ each originate from the ten dimensional metric and antisymmetric tensor
respectively. The remaining $16$ gauge fields may be identified with the
diagonal generators of $SO(32)$ or $E_8\times E_8$.

\vspace{.1in}
For later convenience, we introduce the matrix $M_{ij}$ in the
coset space (3) and rewrite the effective action (1) in the
generic form following the convention of Townsend\cite{9},

\be
S_h= \int d^5x\ {\sqrt {-g}}\ e^{-\P}\big [ R+ (\pr \P )^2
-{1\over12} H^2 -{1\over4}F^{(a)}C_{ab} F^{(b)} - {M}_{ij}\pr
_{\mu} \varphi ^i\pr ^{\mu}\varphi ^j \big ],
\ee
where $C_{ab}$ is a positive definite matrix related to $M_{ij}$ by a
matrix transformations and is a function of the scalars $\varphi ^i$. The
explicit form of the matrix $C_{ab}$ and $M_{ij}$ can be calculated for
the toroidal compactification.

\vspace{.1in}
Now, let us invoke the Poincare string-particle duality in five
dimensions relating the three rank field strength to its
dual $i.e.$ a two rank field strength. We write the duality transformation
as
\be
e^{-\P}\ \pr ^{[\mu}B^{\nu\rho ]} = {1\over{{2!}{\sqrt{-g}}}}
{\e ^{\mu\nu\rho\s\l }} {\tilde F}_{\s\l}\ ,
\ee
where $\tilde F$ is a Poincare dual of the three form field
strength $H$ which is only specific to five dimensions. Rewriting the the
action in eq.(4) in terms of the gauge field strengths only, we arrive at

\bea
S_h= \int d^5x {\sqrt {-g}}e^{-\P}\Big [ R+ (\pr \P )^2
&&-{1\over4} e^{2\P} {\tilde F}^2 -{1\over4}F^{(a)}C_{ab} F^{(b)}
- M_{ij}\pr _{\mu} \varphi ^i\pr ^{\mu}\varphi ^j \Big ]\nonumber\\
&&+{1\over4} \int \ A^{(a)}\w {\cal L}_{ab}\ F^{(b)}\w {\tilde
F}_{\s\l}\ ,
\eea
with a topological term. The last term arises out of the gauge
Chern-Simon term in the definition of three form $H$ as defined in eq.(2).
Note that, there are altogether $27$ gauge fields in eq.(6); out
of which the one corresponding to the dual of the two form potential
appears with a different string coupling from the remaining $26$ of them.
The action in eq.(6) can be given an underlying particle interpretation,
as we proceed, by identifying with a supergravity action in five dimensions.

\section{Eleven dimensional supergravity compactified on $K3\times T^2$}

In this section, we present a five dimensional theory by
dimensional reduction of $N=1$ eleven dimensional supergravity\cite{35}.
We seek for the $D=5$ theory from the present perspective of duality where
it has been conjectured that the strong coupling limit of a ten
dimensional Type $IIA$ string theory is an eleven dimensional
supergravity theory. Thus it is natural to expect that the spectrum of
Bogomol'nyi saturated states in the two theories to be identical in
order to test the conjecture. We start with the conjecture\cite{5,7,9}
that an eleven dimensional supergravity  with an underlying low energy
effective fivebrane theory after double dimensional reduction on $K3$ can
be identified with a low energy effective heterotic string theory
compactified on $T^3$. Now, we further compactify the corresponding
seven dimensional theories with the metric, three form potential and the
gauge fields on a two torus $T^2$ to obtain a five dimensional theory. We
find that the $N=1$, $D=11$ supergravity compactified on $K3\times
T^2$ has $N=2$ supersymmetry in the resulting five dimensions. A similar
type of compactification of an eleven dimensional supergravity theory on
$K3\times T^3$ has been discussed in a different context in ref.\cite{38}.
At present, there is a strong evidence \cite{39} that a ten dimensional
$E_8\times E_8$ heterotic string is related to an eleven dimensional
theory on an orbifold $R^{10}\times S^1/Z_2$, which is the strong coupling
limit of the string theory.

\vspace{.1in}
With this motivation, we consider the bosonic part of the eleven
dimensional $N=1$ supergravity action\cite{35} with an underlying
low-energy membrane world-volume. The massless bosonic field contents are
the metric $G$ and a three form potential ${\cal C }_3$. We write the
action as

\be
S= \int d^{11}x\ {\sqrt {-G}}\ \Big ( {\cal R}^{(11)} +
E_4^2 \Big ) + {1\over{(12)^2}}\int \ {\cal C}_3 \w E_4 \w E_4 \ ,
\ee
where ${\cal R}^{(11)}$ is the eleven dimensional scalar curvature and $E_4
= d{\cal C}_3$ is the four form antisymmetric field strength.

\vspace{.1in}
Now as discussed in ref.\cite{5,9}, the simultaneous dimensional
reduction of the eleven dimensional membrane theory (7) on $K3$ gives
rise to a three form potential ${\cal C}_3$, $22$ abelian gauge fields $\hat
{\ca}$ and a metric ${\cg}$ in seven dimensions and the worldvolume
reduces to a worldsheet. One can write down the supergravity effective
action as

\bea
S=\int d^7x&&{\sqrt {-\cg }}e^{{2\hat\p }\over3}\Big [
R^{(7)} + {1\over3 }(\pr {\hat\p })^2 +{1\over8}
Tr({L'}\pr {K'} L' \pr {K'}) -{1\over48 }
{{\hat E}_4}^2 \Big ] \nonumber\\
&&-{1\over4 }\int d^7x\ {\sqrt {-\cg }}{\hat F}^{(I')}{(L'K'L')}_{I'J'}
{\hat F}^{(J')} + {1\over{(12)^2}}\int \ {\cal C}_3\w {\hat
F}^{I'}\w L_{I'J'}{\hat F}^{J'}
\eea
where $\hat\p$ is the seven dimensional dilaton, ${\hat F}^{(I')}$ for
$I',J'=1,2 \dots ,22$ are the gauge field strengths originated from the
four form field strength, $K'$ corresponds to the moduli field on $K3$ and
${L'}_{I'J'}$ is a constant $(22\times 22)$ matrix defined on $K3$. The four
form field strength in seven dimensions is defined as ${\hat E}_4\ =
d{\hat{\cal C}}_3$. Now further compactifying on a two torus $T^2$
following the prescription in ref.\cite{37}, we get

\bea
S=\int d^5x&&{\sqrt {-\tilde g}}e^{{2\p}\over3} \Big [
R^{(5)} + {1\over3}{\tilde g}^{\mu\nu}\pr _{\mu}\p \pr
_{\nu}{\p } + {1\over8}Tr (\pr _{\mu} G^{\a\b}\pr ^{\mu} G_{\a\b})
+{1\over8} {\tilde g}^{\mu\nu}Tr(L\pr _{\mu}{K} L \pr
_{\nu}{K})\nonumber\\
&&-{1\over4}F^{(\a )} G_{\a\b }F^{(\b )}-{1\over48}
{F'}_{\mu\nu\rho\s} {F'}^{\mu\nu\rho\s} -{1\over{12}}
{{\tilde H}^{(\a)}}_{\mu\nu\rho} {{\tilde
H}_{(\a)}}^{\mu\nu\rho} -{1\over4} {\tilde F}^{(\a\b
)}_{\mu\nu}G_{\a\g}G_{\b\d}{\tilde F}^{(\g\d ){\mu\nu}} \Big ]
\nonumber\\
&&-{1\over4}\int d^5x {\sqrt {-{\tilde g} \ G }}\Big [{F^{(I)}}_{\mu\nu}
(L{K}L)_{IJ}{F^{(J)}}^{\mu\nu} +2 (LKL)_{IJ} \pr_{\mu} {\chi^I}_{\a}\pr ^{\mu}
{\chi^J}^{\a}\Big ]\nonumber\\
&&+{1\over4}\int \Big [L_{IJ}\; A\w F^I\w F^J
+ L_{IJ}\; {\tilde B}\w F^I\w F^J
+ {1\over{3}}L_{IJ}\; C_3\w d\chi ^I\w d\chi ^J \Big ] \ ,
\eea
where $I,J,=1,2,\dots 22$ and $G$ is the deteminant of $G_{\a\b }$.
$(LKL)_{IJ}$ is a $22\times 22$ matrix and is a function of the moduli
fields $A^I\equiv \chi^I$. The five dimensional dilaton $\p$ is originated
from the compactification of the eleven dimensional supergravity theory
and is related to the seven dimensional dilaton $\hat\p$ through field
redefinitions. The background field configurations contain the
Kaluza-Klein gauge fields ${\tilde A}^{(\a )}$ for $\a =1,2$.
Along with that there are one form potentials $A^{(\a )}$, two
form potentials ${\tilde B}^{(\a )}$ and a three form potential
$C_{\mu\nu\rho}$ which have originated from a single three form potential
${\hat{\cal C}}_3$ in seven dimensions. By now it is needless to say that the
three form field strengths and the four form one are defined with
appropriate gauge Chern-Simon terms and are given by

\bea
&{{\tilde H}^{(\a )}}_{\mu\nu\rho} =\pr _{[ \mu}
{B^{(\a)}}_{\nu\rho ]} - {A^{(\a\b )}}_{[ \mu} G_{\b\d}{{\tilde
F}^{(\d )}}_{\nu\rho ]}\nonumber\\
{\rm and}\qquad \qquad\qquad\qquad
& {F'}_{\mu\nu\rho\l} =\pr _{[ \mu}
C_{\nu\rho\l ]} - F^{(\a )}_{[\mu\nu} G_{\a\b}B^{(\b )}_
{\rho\l ]}.
\eea

\noindent
There are altogether $26$ gauge fields, out of which $22$ owe
their origin to $K3$ compactification. Of the remaining four,
two gauge fields appear from the seven dimensional metric when we
compactify on $T^2$ and other two come from the antisymmetric tensor
field. Note that in ref.\cite{40}, various dimensionally reduced field
strengths obtained from the higher dimensional one have also been
discussed in a different context.

\section{Poincare duality}

In this section, we apply Poincare duality on the five dimensional
supergravity theory, discussed in section III, to obtain a dual theory. We
have calculated the background fields in the five dimensional supergravity
theory and find them to be identical in number to that of the low-energy
effective heterotic string compactified on a five torus after using
the Poincare duality on certain field strengths. We explicitly show that with
a Weyl scaling of the metric and twisting the field strengths, it is possible
to identify the $D=11$ supergravity action compactified on $K3\times T^2$
with that of heterotic string action on $T^5$.

\vspace{.1in}
Let us consider the five dimensional action (9) obtained from the eleven
dimensional supergravity compactified on $K3\times T^2$ and rescale
the metric as follows :

\be
{\tilde g}_{\mu\nu}=e^{-2\P}g_{\mu\nu}\;\; and\qquad\p =3\P \ ,
\ee
where $g_{\mu\nu}$ is the new metric, the $\P$ is the dilaton field
after rescaling and subsequently in the later part of this section, we
will identify them with that of the heterotic string effective action in
five dimensions. Now, we write down the action from eq.(9) after the
rescaling of the metric as in eq.(11),

\bea
S=\int d^5x{\sqrt {-g}}&&e^{-\P} \Big [
R^{(5)} + g^{\mu\nu}\pr _{\mu}\P \pr
_{\nu}{\P } +{1\over8} g^{\mu\nu}Tr(L\pr _{\mu}{K} L \pr
_{\nu}{K})+{1\over4} g^{\mu\nu}Tr(\n\pr _{\mu}{M} \n \pr
_{\nu}{M})\nonumber\\
&&-{1\over4}e^{2\P}{F^{(\a )}}_{\mu\nu}\n _{\a\b}F^{(\b)\mu\nu}
-{1\over{12}}e^{4\P} {{\tilde H}^{(\a )}}_{\mu\nu\rho}{\n }_{\a\b}{\tilde
H}^{(\b ){\mu\nu\rho}} -{1\over4} e^{2\P}{\tilde F}_{\mu\nu}{\tilde
F}^{\mu\nu} \nonumber\\
&&-{1\over4}{F^{(I)}}_{\mu\nu}
(L{K}L)_{IJ}{F^{(J)}}^{\mu\nu} -{1\over2}e^{-2\P}
(LKL)_{IJ} \pr_{\mu} {\chi^{I,\a}}\n_{\a\b}\pr ^{\mu}{\chi^{J,\b}}\Big ]
\nonumber\\
&&+{1\over4}\int \Big [ L_{IJ}\ A\w F^I\w F^J + L_{IJ}\ B\w F^I\w
F^J\nonumber\\
&&+{1\over{3}}L_{IJ}\ C_3\w d\chi ^I\w d\chi ^J + 2 \Psi\ {\cdot}\
F^{(\a)}\w {\n}_{\a\b} H^{(\b)} \Big ] \ ,
\eea
where for simplicity, we have taken the matrix
$$G_{\a\b}=\n _{\a\b} =\pmatrix {{1} & {0} \cr {0} & {1} \cr }. $$
The three form antisymmetric field strength ${\tilde H}^{(\a)}$ in eq.(12)
is defined with the gauge Chern-Simon term as mentioned in eq.(10).
The dual of four form field strength $F'$ is a five dimensional pseudo
scalar axion $\Psi $ which can be obtained by using the Poincare duality
transformations given as

\be
{F'}^{\mu\nu\rho\s}=
{{{\e}^{\mu\nu\rho\s\l}}\over{\sqrt{-g}}}\ e^{-5\P}\ \pr _{\l}\Psi \ .
\ee

The two dimensional moduli field $M$ in action (12) is
constructed with the five dimensional axion $\Psi$ and the dilaton
coupling $e^{-\P}$ for consistent counting of the scalar fields
parametrizing the moduli field $\cm $ of the heterotic string on $T^5$.
The $SL(2,R)$ parametrization of the moduli field can be written as
\be
{M}=\pmatrix {{e^{-2\P}} & {\Psi e^{-2\P}} \cr {\Psi e^{-2\P}}
   & {\Psi ^2 e^{-2\P}}+ e^{2\P} \cr }\ ,
\ee
leading to a manifestly $SL(2,R)$ invariant term in the action (12).
However, we have not succeded to write an $SL(2,R)$ invariant form of the
full action (12). Now, we use the string-particle Poincare duality on
three forms ${\tilde H}^{(\a )}$ and a two form field strength ${\tilde
F}$ in order to identify the dual action with that of a heterotic string
discussed in section III. In passing, we would like
to point out an interesting feature of the duality transformation (13) and
the construction of the moduli (14). Notice that the last term in (12) is a
topological term, which involves the two form $F^{(\a )}$, the three form
$H^{(\a )}$ field strengths ($\a =1,2$) and $\Psi $ playing the role of
axion ( normally denoted by $\t $ in four dimensions ). Moreover, it is
easy to see that $M$ parametrizes a moduli $SL(2,R)/SO(2)$ similar to
a dilaton-axion system in four dimensions.

\vspace{.1in}
\noindent The dual of the three form ${\tilde H}^{(\a)\mu\nu\rho}$ in five
dimensions can be written as

\be
{\tilde H}^{(\a )\mu\nu\rho}=
   {{{\e}^{\mu\nu\rho\s\l}}\over{{2!}\sqrt{-g}}}\ e^{-3\P} {{\hat
   F}^{(\a )}}_{\s\l}
\ee
and that of the two form ${\tilde F}^{\mu\nu}$ is
\be
{\tilde F}^{\mu\nu}=
   {{{\e}^{\mu\nu\rho\s\l}}\over{3!{\sqrt{-g}}}}\ e^{-\P}
   {\hat H}_{\rho\s\l} \ .
\ee
Now, we write the $D=11$ supergravity action compactified on $K3\times
T^2$ in terms of dual fields using eqs.(13),(15) and (16) as

\bea
S=\int d^5x&&{\sqrt {-g}}e^{-\P}\Big [R^{(5)} + g^{\mu\nu}\pr _{\mu}\P
\pr {\nu}{\P } +{1\over8} g^{\mu\nu}Tr(L\pr _{\mu}K L \pr_{\nu}K)
+ {1\over4} g^{\mu\nu}Tr(\n\pr _{\mu}M \n \pr_{\nu}M) \nonumber\\
&&-{1\over4}e^{2\P}{F^{(\a)}}_{\mu\nu}\n _{\a\b}F^{(\b )\mu\nu}
-{1\over4}e^{-2\P} {{\hat F}^{(\a )}}_{\mu\nu}{\n }_{\a\b}{\hat
   F}^{(\b ){\mu\nu}} -{1\over{12}} {\hat H}_{\mu\nu\rho}{\hat
   H}^{\mu\nu\rho} \nonumber\\
&&-{1\over4}{F^{(I)}}_{\mu\nu}
(LKL)_{IJ}{F^{(J)}}^{\mu\nu} -{1\over2}e^{-2\P}
   (LKL)_{IJ} \pr_{\mu} {\chi^{I,\a}}\n _{\a\b}\pr ^{\mu} {\chi^{J,\b}}
   \nonumber\\
&&+{1\over2}e^{-2\P} \Psi\ {\cdot}\ {{\hat F}^{(\a )}}_{\mu\nu}{\n
}_{\a\b}F^{(\b ){\mu\nu}} - {1\over2}e^{-4\P}\chi^I \cdot\ L_{IJ}\pr^{\mu}
\chi ^J\pr_{\mu}\Psi -{1\over2}e^{-2\P}\chi^I \cdot\ L_{IJ}
F^{J{\mu\nu}}{\hat F}_{\mu\nu}\Big ] \nonumber\\
&&+{1\over4}\int \Big [ L_{IJ}\ A\w F^I\w F^J
+ A^{(\a )}\n _{\a\b}\w F^{(\b )}\w {\hat F}+
 A^{(\a)}\w {\n}_{\a\b} F^{(\b)} \w {\hat F}\Big ]
\eea

\noindent
There are altogether $26$ gauge fields ${{\ca }_{\mu}}^{(a)}$ for
$a=1,2,\dots 26$, a single antisymmetric potential ${\hat B}_{\mu\nu}$,
whose field strength is ${\hat H}_{\mu\nu\rho}$ and a
set of moduli fields $\chi $'s in the action (17). Suitably incorporating
the $K3$ moduli fields $K$, $SL(2,R)$ invariant moduli $M$
and the $(LKL)_{IJ}$ from $K3\times T^2$, arising out of the compactification
on $K3\times T^2$, we define ${\tilde M}_{ij}$ and ${\tilde C}_{ab}$, to
rewrite the above rescaled action (17) in terms of the redefined moduli
fields $\tilde\cm $. In terms of Einstein metric some of the terms in the
action (17) can be shown to be manifestly $SL(2,R)$ invariant. Note that
$\tilde\cm $ represents a set of $105$ scalars along with the dilaton and
can be identified with that of the heterotic string on a five torus
discussed in the preceeding section. The dilaton and $57$ of scalars
parametrize the moduli $K$ coming from the $K3$ compactification of eleven
dimensional supergravity. The remaining 48 of the scalars arise from the $T^2$
compactification of the seven dimensional supergravity theory, $e.g.$ 44
of them from the gauge field, 3 from metric and the remaining one is
Poincare dual of two form potential. The two form potential ${\hat
B}_{\mu\nu}$ is obtained from the Poincare dual of the spin one field,
which has its origin in the three form potential ${\hat{\cal C}}_3$.
Similarly for the 26 gauge fields in five dimensional supergravity theory;
22 of them come from the gauge fields, two each from the metric and the
two form potential in seven dimensions. We write the respective gauge fields
$A^{(I)}, A^{(\a)}$ and ${\hat A}^{(\a)}$ in a multiplet as

\be
{\cal A}^{(a)}_{\mu\nu}=\ \pmatrix{ {A^{(I)}} \cr { A^{(\a )}_{\mu\nu}} \cr
{{\hat A}^{(\a )}} \cr }
\ee
and write the rescaled supergravity action (17) as

\bea
S=\int d^5x{\sqrt {-g}}e^{-\P}\Big [ R^{(5)}&&+ g^{\mu\nu}\pr
_{\mu}\P \pr _{\nu} \P  - {\tilde M}_{ij}\pr _{\mu} {\varphi} ^{i}\pr
^{\mu}{\varphi }^{j} -{1\over12} g^{\mu\mu '}g^{\nu\nu
'}g^{\rho\rho '} {\hat H}_{\mu\nu\rho} {\hat H}_{\mu '\nu'\rho' }
\nonumber\\
&&-{1\over4}g^{\mu\mu' }g^{\nu\nu '} {\cal F}^{(a)}_{\mu\nu }
{\tilde C}_{ab} {\cal F}^{(b)}_{\mu '\nu '} \Big ]
+{1\over4} \int {\cal L}_{ab}\ A^{(a)}\w {\cal F}^{(b)}\w {\hat
F},
\eea
where

\be
   {\cal L} = \pmatrix{{-L_{IJ}} & {0} & {0}\cr {0} & {\n _{\a\b}} &
   {0} \cr {0} & {0} & {\n _{\a\b}}\cr},
\ee

\be
{\tilde M}_{ij}=\pmatrix {{N_{i'j'}} & {0} &{0} \cr
{0} & {e^{-2\P}\over 2} & {{e^{-4\P}\over 4}L_{IJ}\chi ^J} \cr
{0} & {{e^{-4\P}\over 4}L_{IJ}\chi ^I} & {{e^{-2\P}\over 2}
[(LKL)_{IJ}\times \n_{\a\b}]} \cr }
\ee

and

\be
{\tilde C}_{ab} = \pmatrix { {(LKL)_{IJ}} & {0} &
{e^{-2\P}L_{IJ} \chi ^J} \cr {0} & {e^{2\P}\n _{\a\b}} & {-\Psi
e^{-2\P}\n_{\a\b}} \cr {e^{-2\P}L_{IJ} \chi ^I} & {-\Psi
e^{-2\P}\n_{\a\b}} & {e^{-2\P}\n _{\a\b} } \cr } \ ,
\ee
where $N_{i'j'}$ is a matrix and is a function of the moduli fields
on $K3\times T^2$. The scalar fields ${\varphi }^i$ of the moduli $\tilde
M$ can be identified with those of heterotic string moduli $M$ as in
eq.(6).

\vspace{.1in}
The action in eq.(19) is obtained by compactifying the $D=11$
supergravity theory with an underlying membrane effective action on
$K3\times T^2$. The field content in the massless sector is found to be
equivalent to that of the heterotic string effective action in five
dimensions. Thus a rescaling (11) of the five dimensional supergravity
metric along with the Poincare duality transformations (13), (15) and
(16), correspond to a toroidally compactified heterotic string theory
with $N=2$ space-time supersymmetry. Thus the strong coupling dynamics of
heterotic string theory in five dimensions is identical to the $K3\times
T^2$ compactified $D=11$ supergravity theory.

\section{Strong-Weak coupling duality}

In this section, we discuss the strong-weak coupling
string-particle duality of an eleven dimensional supergravity theory
compactified on $K3\times T^2$, whose action is given in eq.(9).
As discussed in section I, the string-string duality relates
a strongly coupled ten dimensional low energy heterotic string theory
compactified on $T^4$ to a weakly coupled ten dimensional Type $IIA$
string compactified on $K3$ and viceversa. As a result a Type $IIA$ theory
on $K3$ gets enhanced gauge symmetries. Thus it is natural to test
explicitly the strong-weak coupling limit of the compactified eleven
dimensional supergravity theory on $K3 \times T^2$ with that of the Type
$IIA$ string on $K3\times S^1$ inorder to arrive at a set of relations
between various theories.

\vspace{.1in}
\noindent
Now, we start by Weyl scaling the metric $\tilde g$ and redefining the
dilaton field as
\be
{\tilde g}_{\mu\nu}= e^{{2\P '}\over3} {g'}_{\mu\nu}\; ,\;\;\;
\p= -3\P '\ ,
\ee
where the ${g'}_{\mu\nu}$ and $\P '$ denote the Type $IIA$ string metric
and the dilaton respectively. We rewrite the action (9) in terms of these
new fields as

\bea
S=\int d^5x&&{\sqrt {- g'}}e^{-\P'} \Big [R^{(5)} + {g'}^{\mu\nu}\pr
_{\mu}\P ' \pr_{\nu}{\P '} + {1\over4}Tr (\pr _{\mu} G^{\a\b}\pr ^{\mu}
G_{\a\b}) +{1\over8} {g'}^{\mu\nu}Tr(L\pr _{\mu}K L \pr _{\nu}K)\nonumber\\
&&+{1\over4}{g'}^{\mu\nu}Tr (\pr_{\mu}{M}^{-1}\pr _{\nu}{M})
-{1\over4}e^{{-2\P '}\over3}F^{(\a )} G_{\a\b }F^{(\b )}\nonumber\\
&&-{1\over{12}}e^{{-4\P '}\over3}{{\tilde H}^{(\a)}}_{\mu\nu\rho} {{\tilde
H}_{(\a)}}^{\mu\nu\rho} -{1\over4}e^{{-2\P '}\over3} {\tilde F}^{(\a\b
)}_{\mu\nu}G_{\a\g}G_{\b\d}{\tilde F}^{(\g\d ){\mu\nu}} \Big ]\nonumber\\
&&-{1\over4}\int d^5x {\sqrt {-{g'} G }}e^{{4\P '}\over3} \Big [
{F^{(I)}}_{\mu\nu} (LKL)_{IJ}{F^{(J)}}^{\mu\nu} + 2 e^{2\P
'}(LKL)_{IJ} \pr_{\mu} {\chi^I}_{\a}\pr ^{\mu} {\chi^J}^{\a}\Big ] \nonumber\\
&&+ {1\over4}\int \Big [ L_{IJ}\ A\w F^{(I)}\w F^{(J)} + L_{IJ}\ B\w
F^{(I)}\w F^{(J)} + {1\over{3}}L_{IJ}\ C_3\w d\chi^I\w d\chi^J \Big ]
\eea
In this case, the Poincare dual of the four form field strength besides
the gauge Chern-Simon term is given by

\be
{F'}^{\mu\nu\rho\s} ={{{\e}^{\mu\nu\rho\s\l}}\over{{\sqrt{-g'}}}}
\ e^{3\P'} \pr _{\l} \Psi '
\ee

\noindent
We have used the expression in eq.(25) to write the above action (24). The
dual of three form ${\tilde H}$ and the two form $F$ are given by
\bea
&{\tilde H}^{\mu\nu\rho} ={{{\e}^{\mu\nu\rho\s\l}}\over{2!\sqrt{-g'}}}\
e^{{7\P '}\over 3}{\hat F}_{\s\l}\nonumber\\
{\rm and} \qquad\qquad &F^{\mu\nu}
={{{\e}^{\mu\nu\rho\s\l}}\over{3!{\sqrt{-g'}}}} e^{{\P'}\over 3} {\hat
H}_{\rho\s\l},
\eea
where the three form field strength ${\hat H}_{\mu\nu\rho} =
\pr_{[ \mu}B_{\nu\rho ]}$. Now, we write the dual form of the action (24)
using eq.(25)-(26) as

\bea
S=\int d^5x&&{\sqrt {- g'}}e^{-\P'}\Big [
R^{(5)} + {g'}^{\mu\nu}\pr _{\mu}\P ' \pr
_{\nu}{\P '} +{1\over 8} {g'}^{\mu\nu}Tr(\pr _{\mu}K^{-1}\pr
_{\nu}K) +{1\over 4}{g'}^{\mu\nu}Tr (\pr_{\mu}{M}^{-1}\pr
_{\nu}M)\nonumber\\
&&-{1\over 4}e^{{-2\P '}\over3}F^{(\a )} \n_{\a\b }F^{(\b)} +
{1\over 2}e^{{10\P '}\over 3}\ \Psi '\cdot \ F^{(\a )} \n_{\a\b }{\hat
F}^{(\b )} -{1\over 4} e^{{10\P '}\over 3}{{\hat
F}^{(\a )}}_{\mu\nu} \n_{\a\b } {\hat F}^{(\b )\mu\nu}\nonumber\\
&&-{1\over{12}}{\hat H}_{\mu\nu\rho}{\hat H}^{\mu\nu\rho} -
{1\over4}e^{{4\P '}\over 3} {F^{(I)}}_{\mu\nu}
{(LKL)}_{IJ}F^{(J)\mu\nu} -{1\over 2}e^{2\P '}[(LKL)_{IJ}\times
\n _{\a\b }] \pr_{\mu} \chi^{I,\a}{\pr }^{\mu} \chi^{J,\b} \nonumber\\
&&+{1\over 2}\Psi '\cdot \ {\pr }_{\mu }\chi ^I L_{IJ}
{\pr }^{\mu }\chi ^J -{1\over2}e^{{10\P '}\over 3}\chi ^I \cdot \
L_{IJ}F^{J\mu\nu } {\hat F}_{\mu\nu } \Big ] \nonumber\\
&&+{1\over4}\int \Big [ L_{IJ}\ A\w F^{(I)}\w F^{(J)} + \n_{\a\b}
A^{(\a)}\w F^{(\b)}\w {\hat F}_{1} + \n_{\a\b} A^{(\a)}\w F^{(\b)}\w {\hat
F}_{2} \Big ]\ ,
\eea
where we have considered $G_{\a\b}= \n_{\a\b }$ $(\a,\b =1,2 )$ as in
section IV. In order to identify the action (27) with that of a Type $IIA$
string compactified on $K3\times S^1$, we redefine the 26 of the gauge
fields, $A^{(I)}, A^{(\a)}$ and ${\hat A}^{(\b)}$ in a gauge field
multiplet ${\ca}^{(a)}$ for $a=1,2,\dots 26$. As a result, we identify
these gauge fields with that in the Ramond-Ramond sector of a dimensionally
reduced Type $IIA$ string. Identifying the twisted moduli fields on the
$K3\times T^2$ with the matrix $N_{i'j'}$, one can arrive
at the five dimensional string effective action given by

\bea
S=\int d^5x{\sqrt {- g'}}e^{-\P'}\Big [
R^{(5)} &&+ {g'}^{\mu\nu }\pr _{\mu }\P ' \pr
_{\nu}{\P '} -{\cal N}_{ij} \pr _{\mu}{\varphi }^i \pr
^{\mu}{\varphi }^j -{1\over12}{g'}^{\mu\mu '} {g'}^{\nu\nu '}
{g'}^{\rho\rho '} {H'}_{\mu\nu\rho }{H'}_{\mu '\nu '\rho '}\Big ] \nonumber\\
&&-{1\over4}\int d^5x {\sqrt {- g'}}F^{(a)}{\cal D}_{ab}F^{(b)}
-{1\over4}\int L_{ab}\ A\w F^{(a)}\w F^{(b)}\ ,
\eea

\noindent where
\be
{\cal D}_{ab} = \pmatrix { {{e^{{-\p '}\over3}}(LKL)_{IJ}} & {0} &
{e^{{10\p '}\over3}L_{IJ}\chi ^J} \cr {0} & {e^{{-2\p '}\over3}\n_{\a\b}}
& {-\Psi 'e^{{10\p '\over3}}\n_{\a\b}} \cr {e^{{10\p '}\over3}L_{IJ}\chi
^I} & {-\Psi 'e^{{10\p '\over3}}\n_{\a\b}} & {e^{{10\p '}\over3}\n_{\a\b}}
\cr }
\ee
and

\be
{\cal N}_{ij} = \pmatrix { {N_{i'j'}} & {0} & {0} & {0} \cr {0} & {e^{4\p
'}\over2}
& {0} & {0}\cr {0} & {0} & {L_{IJ}\Psi '} & {0} \cr {0} & {0} & {0} &
{{1\over2} e^{2\p '} [(LKL)_{IJ}\times \n _{a\b}]} \cr }
\ee

\vspace{.1in}
\noindent for $a,b = 1,2,\dots 26$ and $i,j =1,2 \dots 105$.
The action in eq.(28) corresponds to a ten
dimensional Type $IIA$ string compactified on $K3\times S^1$. The
transformations (23) along with (25) and (26) acting on the supergravity
action (24) leads to a Type $IIA$ dimensionally reduced string action in
five dimensions with the gauge fields in the Ramond-Ramond sector. In
fact, the five dimensional Type $IIA$ string effective action is obtained
from the corresponding supergravity action by applying the strong-weak
coupling duality as the dilaton of one theory is negative of other (23).
At this point note that the dilaton in the five dimensional
Type $IIA$ string theory has it origin in the toroidal compactification of
the supergravity theory whereas the one in heterotic string has its origin
in the $K3$ moduli fields.

\vspace{.1in}
Thus we have explicitly shown the relation between a dimensionally
reduced supergravity theory with that of heterotic string and the Type
$IIA$ string theory in five dimensions supporting the evidences in ten
dimensions\cite{39}. Since the massless states of a Type $IIA$ string on
$K3\times S^1$ is identical to that of a Type $IIB$ on $K3\times S^1$, we
state a chain of interconnections in five dimensions between various string
theories and the supergravity theory supporting the conjectures in seven
and six dimensions. Furthermore by analyzing the coset
$O(5,21)/[O(5)\times O(21)]$ representing the moduli fields of a Type
$IIB$ string compactified on $K3$\cite{5}, which is indeed similar to
the case of Type $IIB$ compactified on $K3\times S^1$, one can find the
identical massless states in the five dimensional supergravity theory under
discussions. This results in identifying the five dimensional supergravity
theory with that of a Type $IIB$ string in six dimensions which has been
given an interpretation of large radius ($S^1$) limit of the five
dimensional one. We conclude this section by stating that in five
dimensions, the heterotic, Type $IIA$ and Type $IIB$ string theories are
different phases of the dimensionally reduced eleven dimensional
supergravity theory.

\section{Charged solutions of heterotic string effective action}

In support of the dualities in five dimensions as discussed in the
previous sections, we analyze the classical solutions of the heterotic
string effective action and the dual models in this section. We present
magnetically charged solitonic solution of the low energy heterotic
string effective action in five dimensions. The topological (or Noether)
charge of this solution is defined on $S^3$ corresponding to the two form
antisymmetric potential. We extend our analysis for nonsingular point like
string solutions and present a spherically symmetric one of the
string-particle dual effective action with magnetic charge defined on a
two sphere $S^2$. By applying the five dimensional Poincare duality
discussed in section IV, we interprete them as solutions corresponding to
the supergravity theory with an underlying low-energy membrane theory. As
a consequence the BPS solitonic solution of heterotic string theory
remains a solitonic one of the corresponding supergravity theory and
viceversa. However the geometry of the solution changes from $S^3$ to
$S^2$ with magnetic chrage due to gauge field instead of the two form
antisymmetric potential. Furthermore, under the strong-weak coupling
duality, as discussed in the previous section V, the Noether and
topological charges get interchanged\cite{6,7} and the
singular solutions go over to the nonsingular ones and viceversa. We
argue that these solutions describing the geometry with different
topologies are in agreement with the string-particle duality in five
dimensions. The interconnections between different theories in five
dimensions admit the classical solutions of heterotic string theory to be
that of supergravity theory compactified on $K3\times T^2$ and Type $IIA$
theory on $K3\times S^1$.

\vspace{.1in}
Let us start with the ansatz for the classical solution satisfying the
equations of motion of the heterotic string effective action (4). We write
down the classical backgound configurations in five dimensions by
winding of the solitonic string solution of the corresponding six
dimensional one described by Sen\cite{6}.

\vspace{.1in}
\noindent
The string metric defining the invariant distance
\be
ds^2 = \ -dt^2 + e^{\P} \Big (dr^2 + r^2 d{\o _3}^2 \Big )\ ,
\ee
the dilaton $\P $ defining the string coupling
\be
e^{\P} = \ 1 \pm\ {Q\over{r^2}}\ ,
\ee
the three form field strength
\be
H_{\mu\nu\rho}=\ \pm \ {\e }_{\mu\nu\rho\l } \pr ^{\l}\P \; ,
\ee
the gauge fields
\be
A_{\mu} = \ 0\qquad {\rm and\ the\ moduli\ fields}\quad {\varphi ^i}=
const. ,
\ee
where $d{\o_3}^2$ is the $SO(4)$ invariant metric on a three sphere $S^3$.
`$Q$' is the topological charge carried by the solitonic string
states corresponding to the antisymmetric tensor field $B_{\mu\nu}$ in
contrast to that of $A_{\mu}$ which is taken to be zero in this case. The
volume form is defined on a three sphere $S^3$ with $\mu ,\nu
,\rho ,\l = r,\t ,\p ,\psi$. The solution in eq.(31)-(34) satisfy the
background field equations of motion corresponding to the heterotic string
effective action. The Ricci scalar curvature is calculated for the above
backgrounds and found to be nonsingular at $r=0$ and is asymptotically
flat. This magnetically charged solution can be interpreted as a
BPS ``$H$-monopole" like solution in five dimensions which after
compactification of the three sphere to a two sphere becomes a
$H$-monopole solution in four dimensions. The magnetic charge
corresponding to the two form potential in the above eq.(31)-(34) can be
calculated by using the general expression given by
$$ Q\ \equiv \int_{S^3} H .$$

\vspace{.1in}
Now we consider a dual picture in the framework of string-particle
dualities, namely the Poincare duality and the strong-weak coupling
duality, by considering the classical charged solution of the low-energy
heterotic string effective action. We interprete various duals of
the heterotic string effective theory (4) by
replacing the two form potential $B$ with that of a spin one gauge field
$\tilde A$ whose field strength ${\tilde F}$ is defined in eq.(5). In this
framework (6) we seek for spherically symmetric solitonic solution and
discuss their relation with the one in eqs.(31)-(34). The backgrounds
satisfying the equations of motion derived from the action in eq.(6) are :

\vspace{.1in}
\noindent the string metric defining an invariant distance
\be
ds^2 = \ -dt^2 + dx^2 + e^{\P} \Big ( dr^2 + r^2 d{\o _2}^2 \Big )\ ,
\ee
the dilaton $\P$ with a string coupling
\be
e^{\P} = \ 1 \pm\ {{Q'}\over{r}}\ ,
\ee
the only gauge field strength
\be
{\tilde F}_{\mu\nu}=\ \pm {\sqrt 2}\ {\e }_{\mu\nu\l } \pr ^{\l}\P \  ,
\ee
the 26 components gauge field multiplet
\be
A_{\mu}^a=\ 0 \qquad {\rm and\ the\ moduli\ fields}\quad {\varphi ^i}=
const.
\ee
The $Q'$ is the topological charge of the solitonic string states
corresponding to the gauge field $A_{\mu}$. The volume form is defined on
a two sphere $S^2$ with $\mu ,\nu ,\rho = r,\t ,\p $. The only
nonvanishing gauge field satisfying the eq.(37) is given by
$$ {\tilde A}_{\p} =\ {\sqrt 2}Q'\big ( 1-\cos\t \big ) \ .$$

\vspace{.1in}
We have explicitly verified that the above set of metric, dilaton,
gauge fields and the moduli fields satisfy the background field
equations of motion obtained from the dual version of heterotic
string effective action (6) with magnetic charge $Q'$ defined on a two
sphere $S^2$. This represents a magnetically charged BPS solution
with charge defined on a two sphere $S^2$ of radius $r$. In the asymptotic
limit, $r\ra\ \infty$, the scalar curvature vanishes and the solution
becomes flat. The solution (35)-(38) is translationally invariant in
$x-$ direction and represents a spherically symmetric magnetically charged
regular geometry with charge defined on $S^2$ of radius $r$. The solution
can be interpreted as a monopole solution in five dimensions and
by wrapping the $x$-direction in eq.(35) along the string, the solution
can be identified with the $BPS$ monopole solution in four dimensions.
The conserved magnetic charge can also be calculated, with a similar
expression as in the previous section, using the dual field strengths and
can be written as  $$ Q'\ \equiv \int_{S^2} {\tilde F} \ .$$

\vspace{.1in}
The charged solution (35)-(38) can be interpreted as a consequence
of string-particle duality which inturn relates the heterotic
string action to the supergravity, Type $IIA$ and $IIB$ string theories.
The solutions (35)-(38) can be argued to obtain from that of in
eqs.(32)-(34) by applying Poincare duality which transforms a two form
potential to a spin one field. At this point we would like to recall that
a 28 parameter charged classical string solution with nontrivial moduli
field is discussed in ref.\cite{41}.
The consequence of a series of
connections
relating various phases of the eleven dimensional supergravity
compactified on a $K3\times T^2$ to heterotic string compactified on $T^5$
and that of Type $IIA$ string on $K3\times S^1$ in various limits suggests
that the singular solution in a fundamental theory may be interpreted as
the soliton solution of the dual theory and viceversa.

\vspace{.1in}
Note that dual theories admit magnetically charged solutions which are due
to the two form potential and the gauge fields respectively. Since the
Poincare duality relates a two form potential with a spin one gauge field,
we interprete these two charged solutions as dual to each other which
interchanges the geometries namely from a three sphere $S^3$ to a two
sphere $S^2$. However in absence of charges these solutions become
trivial. Further applying strong-weak coupling duality on the heterotic string
effective action (4) which is identified with a supergravity action, one
can arrive at a Type $IIA$ string effective action. As a result, along
with a change in geometries of the magnetically charged solitonic
solutions becomes a singular one and the Noether and topological charges
get interchanged.

\section{Discussion}

In this paper, we have demonstrated the explicit interconnections between
the $N=1$, $D=11$ supergravity compactified on $K3\times T^2$ with that
of a $D=10$ heterotic string theory on a specific $T^5$ and $D=10$ Type
$IIA$, Type $IIB$ string theories on $K3\times S^1$. We have utilized
the string-membrane duality conjecture in seven dimensions along with a
more stronger one, namely string-string duality in six dimensions in order
to provide an evidence for the string-particle duality in five dimensions.
We note that the speciality of five dimensions permit us to construct an
axion which in fact is a Poincare dual to the four form field strength and
appears as a toplogical term in the theory.

\vspace{.1in}
We have also presented magnetically charged classical solutions in five
dimensions with charge defined on a three sphere $S^3$ and a two sphere
$S^2$ respectively. These backgrounds are analyzed and we interprete them
as BPS monopole like solutions analogous to that in four dimensions.
By our construction, we have started with the solutions of different topologies
which are realized as a consequence of the string-particle dualities in
the effective action and manifested in the guise of Poincare duality and
the strong-weak coupling one. These magnetically charged solutions, in fact
can be interpreted as that of the dimensionally reduced supergravity,
Type $IIA$ and $IIB$ string theories through the string-particle dualities
discussed. To be specific, we have presented two solitonic solutions of
the heterotic string effective action on a five torus with the trivial
moduli fields. It is shown that $H$-monopole solutions with charge
corresponding to the two form potential (defined on $S^3$) and that with
respect to the gauge fields (defined on $S^2$) are obtained from the dual
pair of heterotic string theories. In the analysis, we have used the fact
that the two form potential is dual to gauge field and viceversa which is
the origin of string-particle duality in the present context. We note that
the five dimensional duality is accompanied by a change of topology from
$S^3$ to $S^2$ and viceversa in contrast to the four dimensional
particle-particle and six dimensional string-string duality where the
geometries remain unchanged. However in the present context of
nonperturbative duality symmetry, the string-string duality is in much
stronger footing than the string-particle duality in five dimensions and
the string-membrane duality in seven dimensions. In passing, we recall
that the possible puzzle analogous to that of the $H$-monopole problem in
four dimensions\cite{29} still remains unanswared in this context. The
string dynamics in five dimensions has also been analyzed recently in
ref.\cite{42} in a different context.

\bigskip
\bigskip
\bigskip
\bigskip

\noindent {\Large {\bf Acknowledgements}}
\vspace{.2in}

We are thankful to John Schwarz for useful discussions. We would like
to acknowledge valuable discussions with Ashoke Sen and thank him for
carefully reading the manuscript.
One of us (JM) would like to thank E. Witten and the Institute for
Advanced Study for graceous hospitality where a part of this work was done.
This work is partly supported by NSF Grant PHY 92-45317.

\def\np{{\it Nucl. Phys.}\ {\bf B}}
\def\pl{{\it Phys. Lett.}\ {\bf B}}
\def\prd{{\it Phy. Rev.}\ {\bf D}}
\def\prl{{\it Phys. Rev. Lett.}}
\def\ijmp{{\it Int. J. Mod. Phys.}\ {\bf A}}
\def\ml{{\it Mod. Phys. Lett.}\ {\bf A}}
\def\ap{{\it Ann. Phys.}}

\end{document}